# Perfect Anomalous Reflection with an Aggressively Discretized Huygens' Metasurface

Alex M. H. Wong*[1], and George V. Eleftheriades[1]
(1) The Edward S. Rogers Sr. Department of Electrical and Computer Engineering,
University of Toronto, Toronto, Canada, M5S 3G4

## Abstract

This paper investigates the discretization of a periodic metasurface and demonstrates how such a surface can achieve perfect anomalous reflection. Whilst most contemporary theoretical works on metasurfaces deal with continuous current or impedance distributions, we examine how discretization affects a metasurface, and show that in some cases one can discretize a metasurface aggressively – to the extent of having only two cells per spatial period. Such aggressive discretization can lead to great simplifications in metasurface design, and perhaps more surprisingly, a possible performance improvement from continuous metasurfaces. Using this aggressive discretization technique, we report the design of a binary Huygens' metasurface which reflects an incident plane wave at 50° into a reflected direction of –22.5°. Full-wave electromagnetic simulation shows the achievement of anomalous reflection with a power efficiency of 99.1%, which dramatically surpasses the performance of a corresponding passive continuous metasurface, for which the power efficiency is fundamentally limited to 69.6%.

## 1. Introduction

In recent years the metasurface has emerged as a ubiquitous tool for wavefront manipulation. By tuning the electromagnetic parameters of the metasurface, one can modify, almost at will, the amplitude, phase, polarization and direction of reflected and transmitted waves [1, 2]. In particular, Huygens' metasurfaces [3-5] — which feature orthogonal magnetic and electric responses — afford tremendous flexibility in tuning an input plane wave into output waves travelling in any direction of interest, including near-grazing, retroreflection and anomalous directions. A metasurface is typically conceptualized as a surface with continuously varying electromagnetic properties, but it is often implemented as a surface with discretized elements. To avoid discretization effects, very fine discretization steps, such as λ/8 or less, are often used, where λ represents the free space wavelength of the intended illumination.

We think that a study on the discretization of metasurfaces will prove fruitful in the following ways. Firstly, it allows one to take into explicit consideration how discretization affects the metasurface. Secondly, it potentially allows the construction of aggressively discretized metasurfaces, which in some cases feature only two element per grating period. Aggressive discretization may lead to simplified metasurface designs, avoid inter-elemental coupling effects, and lead to metasurface designs which are robust and cost effective. Finally, and perhaps most surprisingly, in some cases one can design aggressively discretized metasurfaces which perform better than their continuous counterparts. In this paper, we shall provide a brief discussion on discretization effects on periodic metasurfaces, and demonstrate the aforementioned advantages with a metasurface which achieves perfect anomalous reflection.

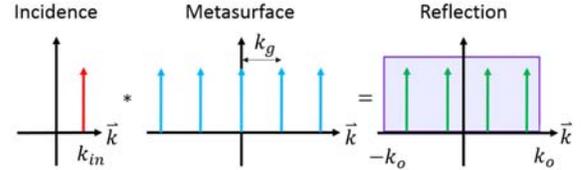

**Figure 1.** A schematic showing the spatial frequency components of an incident plane wave (red), the periodic metasurface (blue), and the resultant reflected waves (green). The region of propagating waves (purple box) is highlighted in the panel for reflected waves. In all three panels, arrows indicate the presence of a plane wave or diffraction order, but do not represent information about their amplitudes and/or phases.

## 2. Metasurface Discretization

As shown in the schematic in Fig. 1, a periodic metasurface diffracts an incoming plane wave into an integer number of propagating diffraction modes. The plane waves and diffraction orders, which are represented in Fig. 1 as $k_y$-space distributions, map straightforwardly into the angular domain through the relationship

$$\sin(\theta) = \frac{k_y}{k_0} \text{ (for } k_y \leq k_0\text{)} \quad (1).$$

For a periodic metasurface with spatial period $\Lambda_g$ and spatial frequency $k_g = 2\pi/\Lambda_g$, the upper-bound on the number of outgoing diffraction modes is

$$N = \left\lceil \frac{2k_0}{k_g} \right\rceil \quad (2),$$

where $\lceil \cdot \rceil$ is the ceiling operator.

A mathematical analysis of discretization effects lies beyond the scope of this paper. Notwithstanding, one can reason that, in order to have sufficient degrees of freedom to determine the strengths and phase of all $N$ diffraction modes, one needs to discretize the metasurface to $N$ elements per spatial period ($\Lambda_g$). From this reasoning, one can deduce that (2) also describes the minimal number of elements per grating period required in the discretization.

We note that in (2) $N$ varies with the inverse of $k_g$. Especially, for $k_g \in [k_0, 2k_0)$, $N = 2$. This means for this range of spatial frequencies, a very aggressive discretization of two cells per grating period suffices for tuning all propagating modes which emerge from the metasurface. In the following we shall apply this to design a metasurface that features perfect anomalous refraction.

## 3. Perfect Anomalous Reflection

Whereas conventional grating theory seems to suggest one can bend a reflected wave into any direction without accruing power loss or requiring gain, recent investigations on metasurfaces have shown otherwise. Upon rigorous analysis involving Maxwell's equations and the corresponding boundary conditions, [6, 7] show that, to achieve total power conversion from an incident plane wave to an arbitrarily directed reflected wave, one needs to realize with a surface which is lossy on half its surface area and active on the other half. The authors of [7] proceed to show such a metasurface [8], which operates by converting an incoming wave into a surface wave in the "lossy" region, and relaunching this surface wave in the "active" region. [9] provides an alternative approach, where auxiliary waves, evanescent in nature, are involved to balance the local power density profile without affecting the far-field radiation, thus avoiding the need for lossy and active regions. Regardless, both methods require complicated designs and manipulation. Without resorting to such methods, the maximum power transfer from an incoming plane wave into a reflected one is limited, by the wave impedance mismatch, to

$$\left|\frac{S_{zr}}{S_{zi}}\right| = \min\left(\left|\frac{\cos\theta_r}{\cos\theta_i}\right|, \left|\frac{\cos\theta_i}{\cos\theta_r}\right|\right) \quad (3),$$

where $S_{zi}$ and $S_{zr}$ are the z-directed Poynting vector of the incident and reflected waves, and $\theta_i$ and $\theta_r$ respectively represent the angles the incoming and reflected waves form with the surface normal.

We shall demonstrate a somewhat surprising fact that, using an aggressively discretized metasurface, one can surpass the aforementioned limit and achieve perfect anomalous reflection using a simple passive metasurface. Our proposed metasurface selectively synthesizes parts on the continuous metasurface which are purely reactive and hence is able to perform anomalous reflection without accruing loss or requiring gain. In reconciling with the point of view of [9], it can also be said that the radical discretization of our metasurface produces evanescent diffraction modes, which can function similar to auxiliary modes to mitigate the need for active or lossy metasurface elements. In the following we shall present target specifications, our design process, and validation results from full-wave simulation using Ansys HFSS.

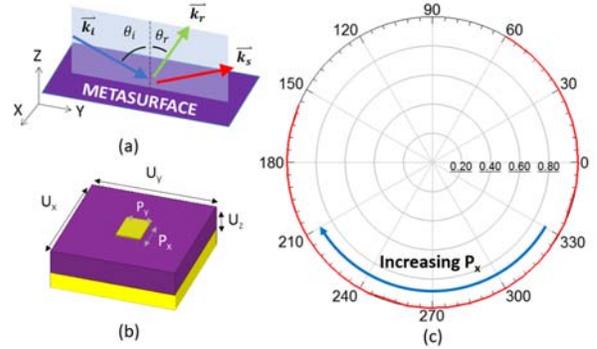

**Figure 2.** Perfect anomalous reflection metasurface design. (a) Diagram showing the geometry of the metasurface, and the wavevectors for the incident ($\vec{k}_i$), anomalously reflected ($\vec{k}_r$) and specular ($\vec{k}_s$) waves. (b) The geometry of a unit cell, with $U_x = U_y = 5.44$ mm, $U_z = 1.575$ mm and $P_y = 0.5$ mm. (c) A plot of the reflection coefficient as $P_x$ is swept from 1 mm to 5 mm.

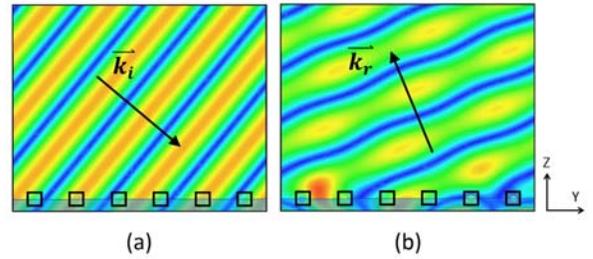

**Figure 3.** Simulation results for the anomalous reflection metasurface. (a) The magnitude of the incident electric field at a reference phase point. (b) The magnitude of the scattered electric field at the same phase point. The designed incidence and reflection directions are drawn in for comparison. The small black squares highlight the dipole locations in this simulation.

## 4. Design and Simulation

The diagram in Fig. 2a sets the geometry of our consideration. The metasurface lies on the xy-plane. A 24 GHz TE-polarized plane wave impinges the metasurface from angle of $\theta_i = 50°$ with respect to the surface normal. We wish to design a metasurface which reflects all the incident power into an outgoing plane wave at $\theta_r = -22.5°$. For these specified angles,

$$k_g = k_0|\sin\theta_r - \sin\theta_i| = 1.15k_0 \quad (4).$$

A substitution into (2) shows that this grating can be aggressively discretized to two cells per grating period; further, it can be found through a stipulation of electromagnetic fields that at these discretization points the local reflection coefficients should differ by 180°.

We implement this metasurface using a ground-backed dipole structure shown in Fig. 2b. We have shown earlier [10-11] that this structure constitutes a Huygens' source, for which the reflection phase can be tuned by changing the dipole length along the direction of the electric field. For this design we employ a substrate with a dielectric constant of $\epsilon_r = 2.2$, a thickness of 1.575 mm, and plated on both sides with perfect conductors. This material system corresponds to that of the Rogers RT/duroid 5880 substrate, with material losses removed. To achieve the required $k_g$ we choose a y-directed unit cell size of $U_y = \pi/k_g = 5.44$ mm. We choose $U_x = U_y$ and $P_y = 0.5$ mm. Fig. 2c shows the reflection coefficient of an infinite 2D array of the unit cell, as a function of a sweep in the dipole length ($P_x$), found from full-wave simulation using Ansys HFSS. As initial dimensions, we choose two lengths to construct our metasurface which give rise to a reflection phase difference of 180°.

Fig. 3 shows full-wave simulation results when we create the metasurface by placing the aforementioned elements side-by-side. In this simulation stage, we re-optimized the dipole lengths to account for reflection phase changes caused by mutual coupling effects with neighbouring elements. The optimized dipole lengths are $P_{x1} = 1.5$ mm and $P_{x2} = 4.71$ mm. A 2D Floquet simulation shows that 99.1% of the power is transferred from 50° incidence to –22.5° reflection, while the remainder of the power remains in the specular direction. This causes the ripple in the scattered wave as seen in Fig. 3b. We can further improve these results with finer optimization steps and improved simulation accuracy. Notwithstanding, our results show that with our aggressively discretized design, we obtained clear improvements over a straightforward, continuous implementation of a passive metasurface, for which the optimal power transfer into the single reflected wave is 69.6% — as found by substituting $\theta_i$ and $\theta_r$ into (3).

## 5. Conclusion

In this work we presented a brief investigation to obtain some measure on the minimal required discretization for a periodic metasurface. Using this knowledge we demonstrated an aggressively discretized perfect anomalous reflection metasurface, which converted an incident wave at 50° into a reflected wave at –22.5°, with a power efficiency of over 99%, as found from full-wave electromagnetic simulation. This represents the simplest yet implementation of a perfect anomalous reflection metasurface. This work also showcases the advantages of aggressively discretizing a metasurface — that they lead to robust metasurfaces designs which can, in some cases, outperform their continuous counterpart.

## 6. References

1. C. L. Holloway et al., "An overview of the theory and applications of metasurfaces: the two-dimensional equivalents of metamaterials," *IEEE Antennas Propag. Mag.*, **54**, 2, April 2012, pp. 10-35.

2. N. Yu, and F. Capasso, "Flat optics with designer metasurfaces," *Nat. Mater.*, **13**, 2, February 2014, pp. 139-150.

3. C. Pfeiffer, and A. Grbic, "Metamaterial Huygens' Surfaces: Tailoring Wave Fronts with Reflectionless Sheets", *Phys. Rev. Lett.*, **110**, 19, May 2013, p. 197401.

4. M. Selvanayagam, and G. V. Eleftheriades, "Discontinuous Electromagnetic Fields Using Orthogonal Electric and Magnetic Currents for Wavefront Manipulation," *Opt. Express*, **21**, 12, June 2013, pp. 14409-14429.

5. A. Epstein, and G. V. Eleftheriades, "Metamaterial Huygens' surfaces: tailoring wave fronts with reflectionless sheets," *J. Opt. Soc. Am. B*, **33**, 2, February 2016, A31-A50.

6. N. M. Estakhri, and A. Alù, "Wave-front transformation with gradient metasurfaces", *Phys. Rev. X*, **6**, 4, October 2016, 041008.

7. V. S. Asadchy, M. Albooyeh, S. N. Tcvetkova, A. Díaz-Rubio, Y. Ra'di, and S. A. Tretyakov, "Perfect control of reflection and refraction using spatially dispersive metasurfaces", *Phys. Rev. B*, **94**, 7, August 2016, 075142.

8. A. Díaz-Rubio, V. S. Asadchy, A. Elsakka, and S. A. Tretyakov, "From the generalized reflection law to the realization of perfect anomalous reflectors," *arXiv Preprint*, arXiv:1609.08041.

9. A. Epstein, and G. V. Eleftheriades, "Synthesis of passive lossless metasurfaces using auxiliary fields for reflectionless beam splitting and perfect reflection" *Phys. Rev. Lett.*, **117**, 25, December 2016, 256103.

10. M. Kim, A. M. H. Wong, and G. V. Eleftheriades, "Optical Huygens' metasurfaces with independent control of the magnitude and phase of the local reflection coefficients", *Phys. Rev. X*, **4**, 4, December 2014, 041042.

11. A. M. H. Wong, P. Christian, and G. V. Eleftheriades, "Binary Huygens' Metasurface: A Simple and Efficient Retroreflector at Near-Grazing Angles", *URSI National Radio Science Meeting*, Jan. 2017, Paper B1-1.